\documentclass[preprint,showpacs,preprintnumbers,amsmath,amssymb]{revtex4}

\usepackage{graphicx}
\usepackage{dcolumn}
\usepackage{bm}

\begin{document}
\draft
\title{Diffraction-free subwavelength-beam optics}
\author{S. V. Kukhlevsky,  M. Mechler}
\address{Institute of Physics, University of Pecs, Ifjusag u. 6, Pecs 7624,
Hungary}
\begin{abstract}
Diffraction is a fundamental property of light propagation. Owing to this phenomenon,
light diffracts out in all directions when it passes through a subwavelength slit.
This imposes a fundamental limit on the transverse size of a light beam at a given
distance from the aperture. We show that a subwavelength-sized beam propagating
without diffractive broadening can be produced in free space by the constructive
interference of multiple beams of a Fresnel source of the respective
high-refraction-index waveguide. Moreover, it is shown that such a source can be
constructed not only for continuous waves, but also for ultra-short (near single-cycle)
pulses. The results theoretically demonstrate the feasibility of
$completely$ $diffraction$-$free$ $subwavelength$-$beam$ $optics$, for both continuous
waves and ultra-short pulses. The approach extends operation of the near-field
subwavelength-beam optics, such as near-field scanning optical microscopy and spectroscopy,
to the "not-too-distant" field regime (0.5 to about 10 wavelengths).

\end{abstract}

\maketitle
\section{Introduction}
Diffraction is one of the fundamental laws of physics. It affects all classical and
quantum mechanical fields without exception. Owing to the law, it is impossible
in quantum mechanics to define at a given time both the position of a matter wave-packet
and its direction to an arbitrary degree of accuracy. Because of the diffraction
phenomenon, light spreads out in all directions after passing a slit smaller than
its wavelength. The harder one tries to decrease the beam transverse dimension by
narrowing the slit, the more it broadens out. Similarly, the beam width dramatically
increases with increasing the distance from the slit. Thus, the diffraction imposes
a fundamental limit on the transverse dimension of a beam at a given distance from
an aperture and consequently limits the resolution capabilities and makes harder
the position requirements of subwavelength-beam optical devices, such as near-field
scanning optical microscopes (NSOM) and spectroscopes (see, for
example \cite{Rayl,Beth,Ash,Betz}).

In order to completely avoid the diffractive broadening of a beam, generally
speaking, one can use the two main approaches. A light beam can be confined to
subwavelength transverse dimensions by multiple total internal reflections at
boundaries of a high-refractive-index waveguide \cite{Marc}. Unfortunately, the guided beam
spreads out in all directions after passing the waveguide output-aperture. Another
approach uses "diffraction-free" beams \cite{Brit,Ziol,Durn,Gori}. A diffraction-less beam, for instance
a Bessel-type beam, propagates in free space without diffractive broadening. Although,
the diffraction-free beams, the widths of which greatly exceed the wavelength, have been
well understood and realized experimentally, neither theoretical principle nor blueprint
of the diffraction-free subwavelength-beam optics has been presented up to now.

In the present article we show that a subwavelength-sized beam propagating
without diffractive broadening can be produced in free space by the constructive
interference of multiple beams of a Fresnel source of the respective
high-refraction-index waveguide. The results theoretically demonstrate the
feasibility of completely diffraction-free subwavelength-beam optics,
for both continuous waves and ultra-short pulses. The approach extends operation of
the near-field subwavelength-beam optics to the "not-too-distant" field regime
(0.5 to about 10 wavelengths).
\section{Theoretical analysis and discussion}
In this section, we show that the diffractive broadening of a subwavelength-sized beam can
be completely avoided. The approach involves a recently established relation between the
waveguide and free-space optics \cite{Kuk1,Cann}. Although, these areas of optics are usually
considered to be independent of each other, it was showed that the fields confined
by a high-refractive-index waveguide, whose width exceeds the
wavelength $\lambda$, could be reproduced in free space by a Fresnel source of this
waveguide. The basic concept of the Fresnel-waveguide source of a diffraction-free
beam is quite simple. The concept is demonstrated in Fig. 1 for a plane-parallel guide
having total-reflection walls. In the approach, the boundaries of the waveguide are
replaced by virtual sources \cite{Kuk1}. The diffraction-free beam $E'(x',z,t)$ confined by
the waveguide is supported in free space at points $(x',z)$ by the constructive
interference of multiple beams $E'_{n}(x',z,t)$ of a Fresnel source of the waveguide:
\begin{eqnarray}
E'(x',z,t)=\sum_{n=-M}^{M}E'_n(x',z,t),
\end{eqnarray}
where the number $2M+1$ of the beams $E'_n(x',z,t)$ depends on their widths
at the distance $z$ from the source; $n=0$,$\pm{1}$,$\pm{2}$,...,$\pm{M}$; $z>0$ and
${\mid}x'{\mid}<a$. The beam $E'_n(x',z,t)$ is
emerged from the $n$-th zone of the Fresnel source having the field distribution
\begin{eqnarray}
E_n(x,0,t)=E_0(x_n,0,t){\exp}(i{\pi}n),
\end{eqnarray}
which is obtained by the periodic $(x_n=x{\pm}2na)$ translation of the field $E_0(x,0,t)$
and the ${\pi}n$-change of its phase; $E_0(x,0,t)$ is the field at the input aperture; $z=0$
and ${\mid}x{\mid}<a$. Thus, the Fresnel-waveguide $\sum_{n=-M}^{M}E'_n(x',z,t)$ is
constructed by the periodic translation and the phase change of the beam $E'_0(x',z,t)$
emerged from the waveguide aperture.

The above-described approach, which was originally developed by
using the Helmholtz-Kirchhoff integral theorem \cite{Kuk1}, fails
when the waveguide width $2a$ is close to the wavelength
$\lambda$. It is surprising that in the general form of Eqs. 1 and
2, as it will be shown below, the approach provides solution of
the problem also in the case of subwavelength waveguides (see,
Figs. 2-4). Figure 2 presents the normalized energy flux
distribution
$S_z^{norm}=(c/8{\pi})Re(\vec{E}{\times}{\vec{H}^*})_z$ of a
diffraction-less beam computed for the Fresnel source of a
subwavelength ($a = 0.05\lambda$) waveguide, at the three
distances $z$ from the source. In the figure the normalized flux
of a single beam, which was used in the construction of the
Fresnel source, is shown at $z=6a$ for comparison. The Fresnel
waveguide $E'(x',z,t)$ was constructed by the translation of the
single beam $E'_0(x',z,t)$ and the periodical change of its phase
(see, Eqs. 1, 2 and Fig. 1). The single beam $E'_0(x',z,t)$ is
formatted by transmission of a plane monochromatic wave through a
subwavelength waveguide (thick slit) with perfectly conducting
walls. The details of the computations of the electric
$\vec{E}=(E_x,0,E_z)$ and magnetic $\vec{H}=(0,H_y,0)$ field
distributions of the transmitted beam are presented in Refs.
\cite{Bet2,Kuk2}. The full width at half maximum (FWHM) and the
energy flux of the diffraction-less subwavelength beam versus the
distance $z$ from the Fresnel-waveguide source are shown in Fig.
3, in comparison to that of the single beam produced by the slit.
The results presented in Figs. 2 and 3 demonstrate theoretically
the feasibility of completely diffraction-free subwavelength-beam
optics. Notice, that at the distance $z=a$, which is generally
accepted for practical near-field scanning optical microscopy
\cite{Bet2}, FWHM of the diffraction-free beam is about two times
lower than that of the single beam. This should provide two-times
increase of the spatial resolution of the near-field scanning
optical microscopy and spectroscopy. Although, the
Fresnel-waveguide source producing a subwavelength
diffraction-free continuous wave is already an unexpected finding,
an example presented in Fig. 4 demonstrates that such a source can
be constructed also for the ultra-short (near single-cycle)
pulses. The figure shows the computed electric field $E_x$ of the
diffraction-free subwavelength pulse formatted by the
Fresnel-waveguide source at the distance $z=3a$. The pulse used
for the construction of the Fresnel waveguide is formatted by
transmission of the femtosecond (near single-cycle) pulse through
the slit \cite{Kuk2}. It should be noted that the diffraction-free
pulse (Fig. 4) is not in conflict with the recently established
uncertainty relation between spatial and temporal uncertainties of
a wave-packet \cite{Kuk3} because of the multiple-beam nature of
the Fresnel-waveguide source. The diffraction-free pulse is
localized in space and time at the expense of increasing the
number of beams (pulses) that support the diffraction-free
wave-packet under its free-space propagation.

The results theoretically demonstrate the feasibility of
$completely$ $diffraction$-$free$ $subwavelength$-$beam$ $optics$,
for both continuous waves and ultra-short pulses. The approach
extends operation of the near-field subwavelength-beam optics to
the far-field regime ($z>0.5{\lambda}$). Notice, in this
connection, that the number of the Fresnel beams supporting the
diffraction-free beam increases and the beam intensity decreases
with increasing the distance $z$ (see, Figs. 1-3). Therefore, the
effective operation of the optics is restricted to the
"not-too-distant" field region (0.5 to about 10 wavelengths). The
optics could improve spatial and temporal resolution capabilities
and positioning requirements of the near-field scanning optical
microscopy and spectroscopy. The approach could also be used for
many other subwavelength-photonic purposes, such as sensors,
communications, optical switching devices and microsources. It
should be noted that the method can be extended to the
3-dimensional beams. In this case, one could use the Fresnel
sources of 3-dimensional subwavelength waveguides \cite{Kuk4}. The
concept of the Fresnel sources of subwavelength waveguides helps
us to understand the studies \cite{Leze,Mart}, which have
demonstrated that a series of parallel grooves surrounding a
nanometre-sized slit in a metal film produces a micrometer-size
beam that spreads to an angle of only few degrees. Another
relevant result is achievement of about 20 percent reduction of
far-field diffraction by structured apertures \cite{Doga}. The
evident parallelism between mechanisms of the formation of a
subwavelength waveguide in free space by the Fresnel multiple-beam
source and the enhancement of light transmission by subwavelength
aperture arrays \cite{Ebbe} should also be noted. The presented
theoretical principle of the Fresnel-waveguide source of a
diffraction-free subwavelength beam could be realized
experimentally using the techniques \cite{Leze,Mart,Doga}, the air
and dielectric guide bends or the hybrid hetero structures (for
example, see the study \cite{Chut} and references therein).

\section{Conclusion}
We have showed that a subwavelength-sized beam propagating without diffractive
broadening can be produced in free space by the constructive interference of multiple
beams of a Fresnel source of the respective high-refraction-index waveguide. The
presented results theoretically demonstrate the feasibility of
$completely$ $diffraction$-$free$ $subwavelength$-$beam$ $optics$, for both continuous
waves and ultra-short pulses. The approach extends operation of the near-field
subwavelength-beam optics to the "not-too-distant" field regime (0.5 to about 10
wavelengths).
\begin{acknowledgments}
This study was supported by the Fifth Framework of the
European Commission (Financial support from the EC for shared-cost
RTD actions: research and technological development projects,
demonstration projects and combined projects. Contract
NG6RD-CT-2001-00602). The authors thank the Computing Services
Centre, Faculty of Science, University of Pecs, for providing
computational resources.
\end{acknowledgments}
\newpage
\begin{figure}
\includegraphics{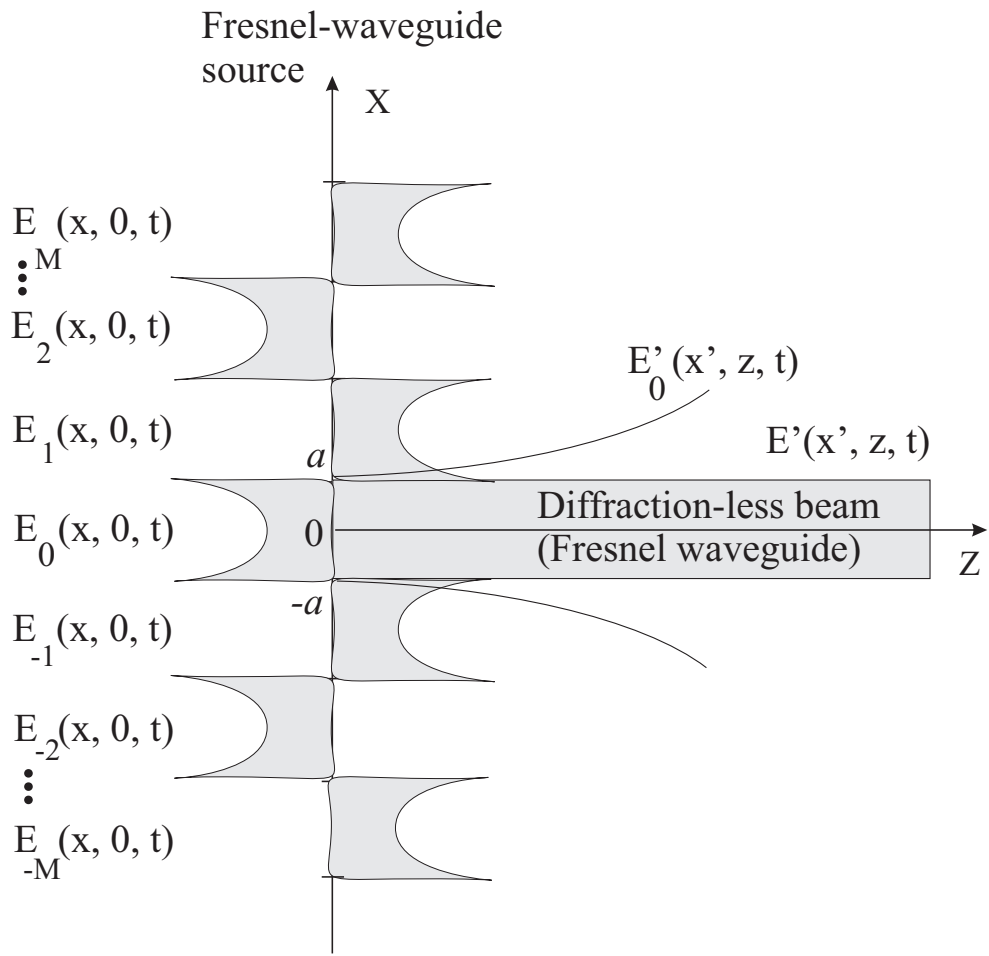}
\caption{\label{fig:epsart}  The construction of a Fresnel source for a plane-parallel
waveguide having total-reflection walls.}
\end{figure}
\newpage
\begin{figure}
\includegraphics{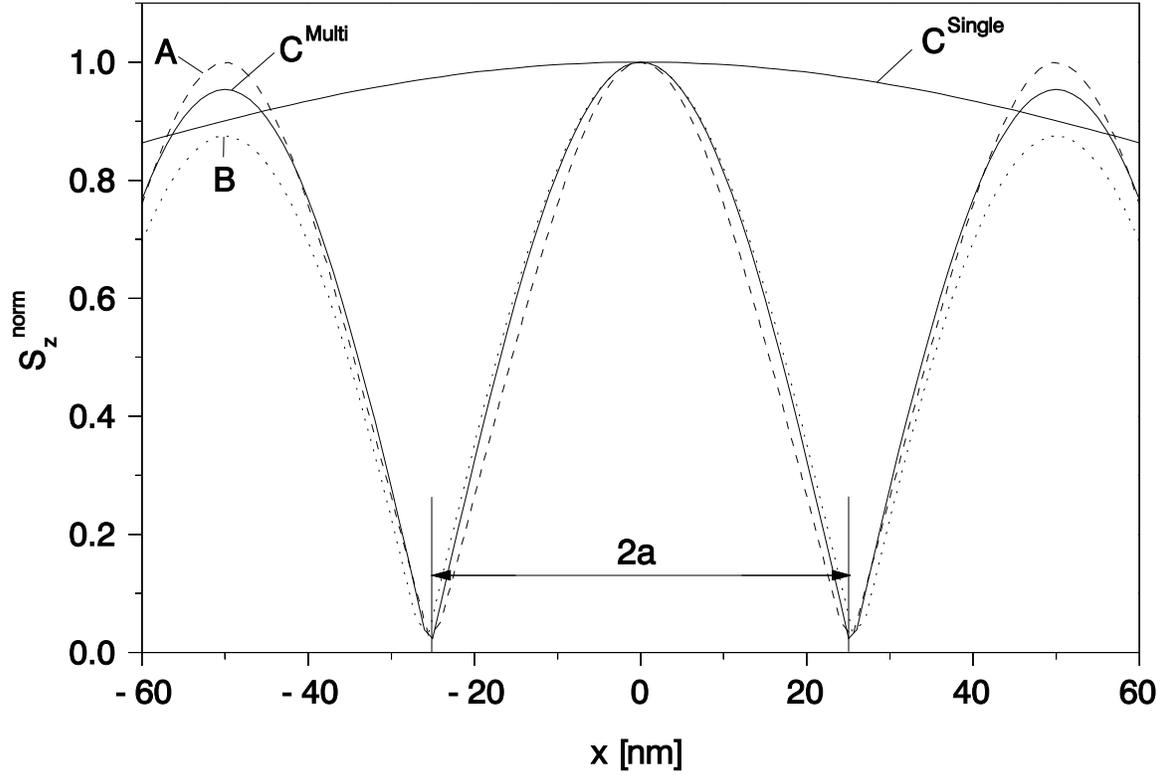}
\caption{\label{fig:epsart}  The normalized energy flux distribution $S_z^{norm}(x')$ of a
diffraction-free subwavelength beam $E'(x',z,t)$ at different distances $z$ from the
Fresnel source of the waveguide: A - $a$, B - $3a$, and C$^{Multi}$ - $6a$. A single beam
$E'_0(x',z,t)$ used for construction of the source is produced by a 50-nm slit (waveguide)
in a perfectly conducting screen of thickness $b=50nm$. The normalized energy flux
distribution C$^{Single}$ of this beam is shown at $z=6a$ for comparison. The number $2M+1$ of
used beams for A and B is 51, for C$^{Multi}$ is 501; $a = 0.05\lambda$ and $\lambda=500nm$.}

\end{figure}
\newpage
\begin{figure}
\includegraphics{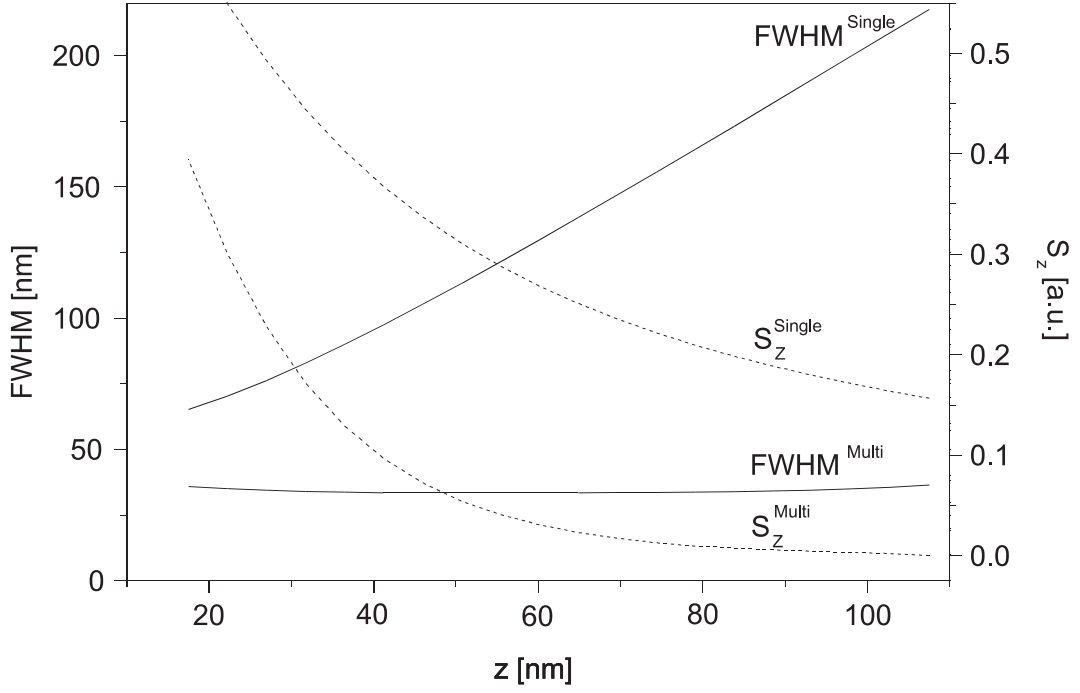}
\caption{\label{fig:epsart} Full width at half maximum (FWHM)$^{Multi}$ and energy flux
$S_z^{Multi}(x'=0,z)$ of a diffraction-less subwavelength beam $E'(x',z,t)$ versus
distance $z$ from the Fresnel source of the waveguide. The width and flux of a single
beam $E'_0(x',z,t)$ used for construction of the source are shown for comparison. Here,
the waveguide parameters are the same as indicated in Fig. 2. The number $2M+1$ of used
beams is 501; $a=0.05\lambda$ and $\lambda=500nm$.}
\end{figure}
\newpage
\begin{figure}
\includegraphics{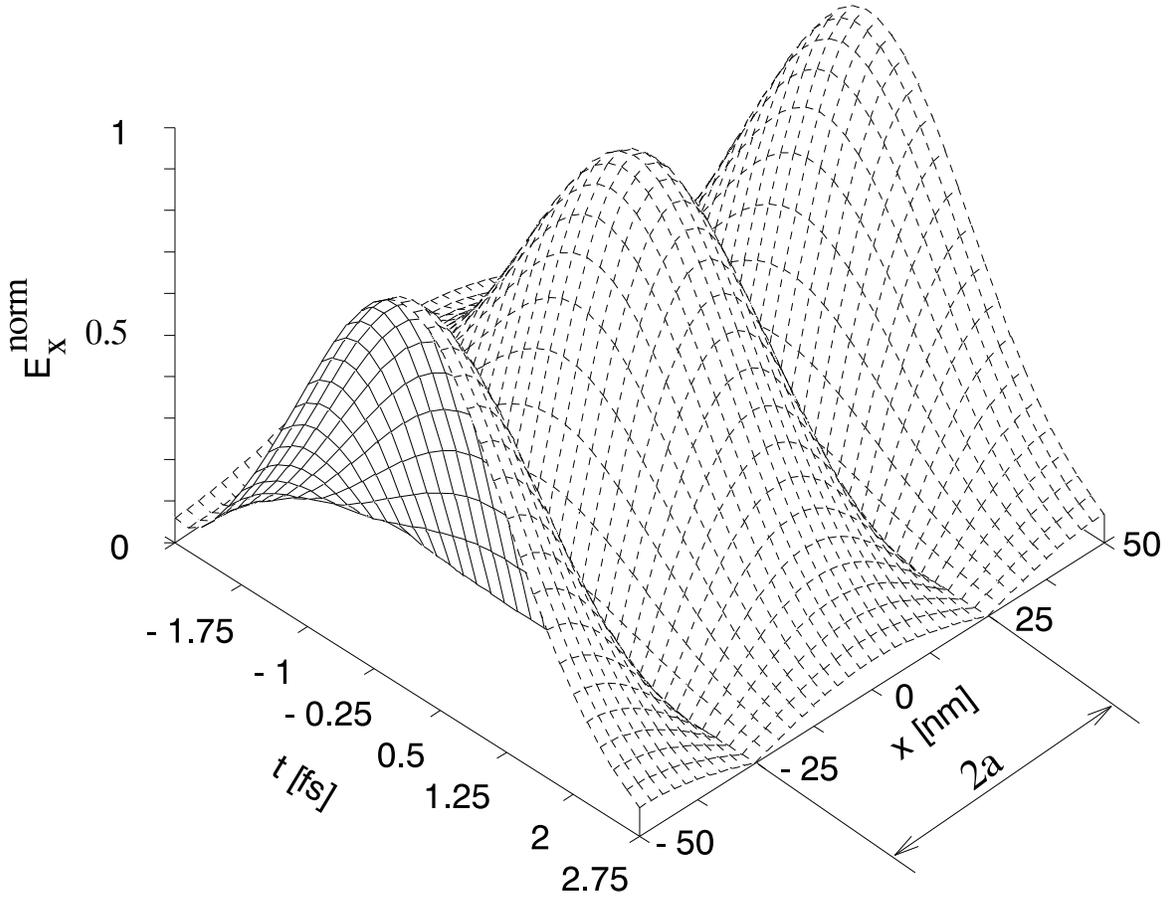}
\caption{\label{fig:epsart} The computed electric field $E_x$ of the diffraction-less
femtosecond (near single-cycle) pulse $E'(x',z,t)$ produced by the Fresnel source of
the subwavelength waveguide at the distance $z=3a$. The parameters of the waveguide
are the same as indicated in Fig. 2. A single pulse $E'_0(x',z,t)$ used for construction
of the source has the pulse length ${\tau}=2fs$ and  central wavelength of the
wave-packet $\lambda_0=500nm$. The number of beams (pulses) is $2M+1 = 18$;
$a=0.05\lambda_0$.}
\end{figure}
\end{document}